\documentclass[prd,aps]{revtex4}
\newcommand{\p}[4]{#1^{#3}_{#2,\,#4}}

\begin{document}

\title{Uniform discretizations: a new approach
 for the quantization of totally constrained systems}

\author{ Miguel Campiglia$^{1}$, Cayetano Di Bartolo$^{2}$ 
Rodolfo Gambini$^{1}$,
Jorge Pullin$^{3}$}
\affiliation {
1. Instituto de F\'{\i}sica, Facultad de Ciencias,
Igu\'a 4225, esq. Mataojo, Montevideo, Uruguay. \\
2. Departamento de F\'{\i}sica,
Universidad Sim\'on Bol\'{\i}var, Aptdo. 89000, Caracas 1080-A,
Venezuela.\\
3. Department of Physics and Astronomy, Louisiana State University,
Baton Rouge, LA 70803-4001}

\begin{abstract}
We discuss in detail the uniform discretization approach to the
quantization of totally constrained theories.  This approach allows to
construct the continuum theory of interest as a well defined, controlled,
limit of well behaved discrete theories.  We work out several finite
dimensional examples that exhibit behaviors expected to be of
importance in the quantization of gravity. We also work out the case
of BF theory. At the time of quantization, one can take two points of
view.  The technique can be used to define, upon taking the continuum
limit, the space of physical states of the continuum constrained
theory of interest. In particular we show in models that it agrees
with the group averaging procedure when the latter exists. The
technique can also be used to compute, at the discrete level,
conditional probabilities and the introduction of a relational
time. Upon taking the continuum limit one can show that one reproduces
results obtained by the use of evolving constants, and therefore
recover all physical predictions of the continuum theory.  This second
point of view can also be used as a paradigm to deal with cases where
the continuum limit does not exist. There one would have discrete
theories that at least at certain scales reproduce the semiclassical
properties of the theory of interest. In this way the approach can be
viewed as a generalization of the Dirac quantization procedure that
can handle situations where the latter fails.
\end{abstract}

\maketitle
\section{Introduction}

Loop quantum gravity has emerged in recent years as one of the major
candidates for a theory of quantum gravity. See \cite{lqg} for recent
reviews. The theory has a mathematically rigorous basis for its
quantum kinematics \cite{kinematics}, which has also been proven to
be unique up to certain assumptions \cite{lost}. 
It also has achieved attractive
physical results in the context of providing a detailed microscopic
picture of black hole entropy and a detailed picture of the big bang
in the context of homogeneous quantum cosmologies.  However, the
problem of the dynamics of the full theory has remained unsettled.
Related to this is the lack of a good understanding of how general
relativity arises as a semiclassical limit of the quantum theory and
the fact that the algebra of quantum constraints ---though free of
anomalies--- does not mimic the algebra of the classical constraints
completely.

The reason for the elusive nature of the dynamics has a well defined
but technical explanation. In the accepted and largely unique
kinematical setup in terms of spin networks 
there is a natural action of the diffeomorphism group 
that is a unitary representation and
therefore the spatial diffeomorphisms are represented without
anomalies. However, the action is not weakly continuous. This means
that the infinitesimal generators of the diffeomorphism group, that
is, the associated Lie algebra, cannot be defined on the Hilbert
space.  In contrast, versions of the Hamiltonian constraint {\em can}
be defined on the Hilbert space.  However, since the classical Poisson
algebra of constraints involves the infinitesimal generators of
diffeomorphisms, it should not be surprising that the classical
constraint algebra is not appropriately reflected in the quantum
constraint algebra. A clear discussion of this problem is present in
the introduction to the recent paper by Giesel and Thiemann
\cite{aqg}. There does not appear to exist an obvious solution to this
problem in the sense of constructing a ``better'' Hamiltonian
constraint operator, for obvious reasons. This obstruction has led
several researchers to consider alternatives to the usual Dirac
approach to the problem of the dynamics. One of the alternatives is
the ``master constraint'' project of Thiemann and collaborators
\cite{phoenix} motivated by 
an earlier proposal by Klauder \cite{klauder}.  Others
consider the covariant ``spin foam'' approach as an alternative, since
one may bypass the construction of the canonical algebra entirely, at
least in some settings. Another approach is the one furthered in this
paper, which we will refer to as ``uniform discretizations''.

Over the last few years we have developed a new paradigm for dealing
with totally constrained theories (see
\cite{paradigm,oriti,ashtekar} for recent reviews).  The approach
has been called ``consistent discretizations'' and a particularly
promising version of it recently introduced \cite{uniform} is called
``uniform discretizations''. To describe the paradigm in a nutshell
one can say that its relationship to totally constrained theories is
the same that lattice QCD has with continuum QCD. The new approach
consists in building a set of discrete theories and to define a
continuum theory as a suitably well defined limit.  Just like one can
view lattice QCD as a tool to define continuum QCD non-perturbatively
one can view the uniform discretization as a tool to define continuum
theories that are totally parameterized, as general
relativity. Therefore the goal is to have a discretized version of the
theory that is well defined, under control and quantizable without
difficulties and only then study how to take the continuum
limit. Other approaches to the dynamics of quantum gravity currently
being considered also use discretizations, but the continuum limit is
immediately taken since the discrete theories are in themselves
inconsistent, i.e. they do not exist.

The naive application of a discretization ---like the ones used in
lattice QCD--- to gravity or other totally constrained field theories
immediately fails since the introduction of the lattice conflicts with
diffeomorphism invariance. From an operational viewpoint, this
translates itself in the failure of constraint algebras to close. In
other words, the discretization of a totally constrained theory
generically yields a discrete theory that is {\em inconsistent}, its
equations cannot be simultaneously solved. This is manifested by the
lack of closure of constraint algebras (more precisely, the
constraints do not structure themselves into an algebra). A typical
example of such an approach is the one followed in numerical
relativity in the Cauchy approach, where it is well known that one
cannot solve simultaneously the evolution equations and the constraint
equations. In that context the usual approach is to ignore some of the
equations ---typically the constraints--- in the hope that if one
starts with data that satisfies them initially, they will remain small
upon evolution. The logic is an a-posteriori one, that is, if one
carries out a simulation and the constraints remain small, then it is
an acceptable approximation to the continuum theory. But it frequently
happens that they do not remain small and the simulation then has to
be abandoned.  Much more serious is the situation if one is interested
in quantizing. To use a set of inconsistent theories to define via a
limiting process a continuum theory at the quantum level is clearly
not logically tenable; to exist a limit one first has to have a
sequence. In quantum mechanics one cannot just pick and choose some
initial data and hope that ``constraints will remain small''. Since
the inconsistent theories do not exist, their quantization makes no
sense and one cannot construct a sensible continuum limit.

This is the reason why we have been studying the possibility of
constructing discretizations of totally covariant theories that are
{\em consistent}, that is, the equations of the discrete theory can be
solved simultaneously. How can this be achieved? Largely by turning
the quantities that in the continuum play the role of Lagrange
multipliers into dynamical variables. The extra variables allow to
make compatible the set of equations that were formerly incompatible.
For instance, in general relativity written in terms of metric
variables in the traditional $3+1$ decomposition, one has twelve
canonical variables and twelve evolution equations, plus four
constraint equations. One has therefore 16 equations for 12 
unknowns. In the continuum the equations are compatible and can
be solved. When discretized however, one ends up with a set of
incompatible equations. If we now consider the three components
of the shift and the lapse as dynamical variables we have 16 
equations with 16 dynamical variables as unknowns and simultaneous
solution of the equations is generically possible.

An immediate question that arises is: if the Lagrange multipliers are
now determined, is this not equivalent to choosing a gauge?  The
answer is obviously negative, since the resulting theory has more
degrees of freedom than the one we started from, it cannot simply be a
gauge fixing of it. The coordinate freedom is indeed encoded in the
extra degrees of freedom of the discrete theory. Still, since the
Lagrange multipliers are determined (and in the case of a totally
constrained theory this means the lapse is determined), how is the
continuum limit to be materialized? The answer is that one can
approximate the continuum theory by choosing initial data in the
discrete theory in a careful way. To put it differently there exist
initial data for the discrete theory such that when evolved the
resulting solution approximates well the solutions of the continuum
theory. There also exist initial data for the discrete theory that do
not approximate the continuum {\em at all}. The continuum limit
therefore is defined by choosing suitably limits within families of
initial data. To do this in a consistent way within a canonical
framework and throughout all of phase space is quite non-trivial. This
is what the ``uniform discretizations'' achieve. There one discretizes
the theory in a way that certain quantities are conserved upon
evolution and therefore one has a well defined way to characterize the
continuum limit to be taken in terms of the initial data.

The uniform discretizations were introduced briefly in reference 
\cite{uniform}. 
The purpose of this paper is to apply the uniform discretization
technique to a set of examples of increasing level of challenge.  Most
of the examples are the same ones as Thiemann and collaborators have
considered in their master constraint program. The organization of the
paper is as follows: in section 2 we briefly review the consistent
and uniform discretization techniques, in section 3 we consider the
first example: homogeneous cosmologies; in section 4 we consider
a model with Abelian constraints, in section 5 a model with 
non-Abelian constraints that structure themselves into a Lie 
algebra, in section 6 we consider a system with a non-Abelian 
and non-compact Lie algebra of constraints, in section 7 
we explore BF theory as a first example of a field theory. We
end with a discussion. The examples of section 4, 5, 6 have
been discussed in detail with the ``master constraint'' approach
\cite{phoenix}.

\section{Consistent and uniform discretizations: a brief review}

\subsection{Consistent discretizations}

We start by considering a continuum theory representing a mechanical
system.  Its Lagrangian will be denoted by ${\hat L(q^a,\dot{q}^a)}$,
$a=1\ldots M$.  This setting is general enough to accommodate, for
instance, totally constrained systems. In such case $\dot{q}$ will be
the derivative of the canonical variables with respect to the
evolution parameter. It is also general enough to include the systems
that result from formulating on a discrete space-time lattice a
continuum field theory, like general relativity.

We discretize the evolution parameter in intervals (possibly varying
upon evolution) $t_{n+1}-t_n=\epsilon_n$ and we label the generalized
coordinates evaluated at $t_n$ as $q_n$. We define the discretized
Lagrangian as

\begin{equation} 
L(n,n+1) \equiv L(q^a_n,q^a_{n+1}) \equiv \epsilon_n{\hat
L}(q^a, \dot q^a)\label{disc} \end{equation}
where
\begin{equation} q^a=q^a_n \quad \hbox{and} \quad \dot{q}^a \equiv
{{q^a_{n+1}-q^a_n}\over \epsilon_n}. \end{equation}

Of course, one could have chosen to discretize things in a different
fashion, for instance using a different approximation for the
derivative, or by choosing to write the continuum Lagrangian in terms
of different variables. The resulting discrete theories generically
will be different and will approximate the continuum theory in
different ways. However, given a discrete theory, the treatment we
outline  is unique. We will take advantage of this
freedom in how to discretize the theory to propose the ``uniform 
discretizations'' in the next section.

The action can then be written as
\begin{equation}
S=\sum_{n=0}^N L(n,n+1).
\end{equation}

It should be noted that in previous treatments we have sometimes
written the Lagrangian in first order form, i.e. $L=\int dt
\left(p\dot{q}-H(p,q)\right)$. It should be emphasized that this is
contained as a particular case in the treatment we are presenting
here. In this case one takes both $q$ and $p$ to be
configuration variables, and one is faced with a Lagrangian that
involves $q_n,p_n$ and $q_{n+1}$ as variables and does not depend on
$p_{n+1}$. The reason we frequently resort to first order formulations
in the various concrete examples we discuss is that the Ashtekar
formulation used in loop quantum gravity is naturally a first order
one and we usually tend to frame things in a closely related way. But
again, there is no obstruction to using either first or second order
formulations with our framework, they are both contained as particular
cases.

If the continuum theory is invariant under
reparameterizations of the evolution parameter, one can show that the
information about the intervals $\epsilon_n$ may be absorbed in the
Lagrange multipliers. In the case of standard mechanical systems it is
simpler to use  an invariant interval $\epsilon_n=\epsilon$.  

The Lagrange equations of motion are obtained by requiring the
action to be stationary under variations of the configuration
variables $q^a$ fixed at the endpoints of the evolution
interval $n=0,n=N+1$,
\begin{equation}
{\partial L(n,n+1) \over \partial  q^a_{n}}+{\partial L(n-1,n)
\over \partial q^a_{n}}=0. \label{lagra}
\end{equation}

We introduce the following definition of the canonically conjugate
momenta of the configuration variables,
\begin{eqnarray}
 p^a_{n+1} &\equiv& {\partial L(n,n+1) \over \partial  q^a_{n+1}}
\label{1}\\
 p^a_{n} &\equiv& {\partial L(n-1,n) \over \partial  q^a_{n}}=-
{\partial L(n,n+1) \over \partial  q^a_{n}}\label{2}
\end{eqnarray}

Where we have used Eq. (\ref{lagra}). The equations (\ref{1}) and
(\ref{2}) define a canonical transformation for the variables
$q_n,p_n$ to $q_{n+1},p_{n+1}$ with a the type 1 generating function
$F_1= -L( q^a_n, q^a_{n+1})$. Notice that the evolution scheme is
implicit, one can use the bottom equation (since we are in the
non-singular case) to give an expression for $q_{n+1}$ in terms of
$q_n,p_n$, which in turn can be substituted in the top equation to get
an equation for $p_{n+1}$ purely in terms of $q_n,p_n$.

It should be noted that there are several other possible choices, when
going from the set of equations (\ref{1},\ref{2}) to an explicit
evolution scheme (see Di Bartolo {\it et al.} \cite{DiBartolo:2004cg}
for further details.)

The canonical transformation we introduced becomes singular as an
evolution scheme if $\displaystyle\left|{\partial^2 L(n,n+1) \over
\partial q^a_{n+1}\partial q^b_{n}}\right|$ vanishes. If the rank of
the matrix of second partial derivatives is $K$ the system will have
$2(M-K)$ constraints of the form,
\begin{eqnarray}
\Phi_A( q^a_n, p^a_n)&=&0\\
\Psi_A( q^a_{n+1}, p^a_{n+1})&=&0.
\end{eqnarray}
And these constraints need to be enforced during evolution, which may
lead to new constraints. We refer the reader for the 
detailed Dirac analysis to Di Bartolo {\it et al.} \cite{DiBartolo:2004cg}. 

To clarify ideas, let us consider an example. 
The model consists of a parameterized free
particle in a two dimensional space-time under the influence of a
linear potential. The discrete Lagrangian is given by,
\begin{eqnarray}
L_n&\equiv&
L(q_n^a,\pi_n^a,N_n,q_{n+1}^a,\pi_{n+1}^a,N_{n+1})\label{la}\\
&=&\pi_n^a(q_{n+1}^a-q^a_n)-N_n[\pi_n^0+ \frac{1}{2}(\pi_n^1)^2+\alpha
q_n^1].\nonumber
\end{eqnarray}
We have chosen a first order formulation for the particle (otherwise
there are no constraints and the example is trivial). However,
this Lagrangian is of the type we considered in this presentation, one 
simply needs to consider all variables, $q^a,\pi^a,N$ as configuration
variables. The system is clearly  singular since the $\pi's$ and
$N$ only appear at level $n$ (or in the continuum Lagrangian, their
time derivatives are absent). When considered as a Type I
generating function, the above Lagrangian leads to the equations
\begin{eqnarray}
\p{p}{\pi}{a}{n+1} &=& \frac{\partial L_n}{\partial \pi^a_{n+1}}
=0, \label{evol11}
\\
\p{p}{q}{a}{n+1} &=& \frac{\partial L_n}{\partial q^a_{n+1}}
=\pi_n^a, \label{evol12}
\\
\p{p}{N}{}{n+1} &=& \frac{\partial L_n}{\partial N_{n+1}} =0,
\label{evol3}
\end{eqnarray}
and
\begin{eqnarray}
\p{p}{\pi}{a}{n} &=& -\frac{\partial L_n}{\partial \pi^a_n}
=-(q_{n+1}^a-q_n^a)+ \pi_n^1 N_n \delta^a_1+ N_n \delta^a_0,
\label{evo21}
\\
\p{p}{q}{a}{n} &=& -\frac{\partial L_n}{\partial q^a_n}
=\pi_n^a+\delta^a_1\alpha N_n, \label{evo22}
\\
\p{p}{N}{}{n} &=& -\frac{\partial L_n}{\partial N_n}
=\pi_n^0+\frac{1}{2}(\pi_n^1)^2+\alpha q_n^1 \label{evo23}.
\end{eqnarray}
The constraints (\ref{evol11},\ref{evol3},\ref{evo22},\ref{evo23}) can
be imposed strongly to eliminate the $\pi$'s and the $N$'s and obtain
an explicit evolution scheme for the $q$'s and the $p_q$'s,
\begin{eqnarray}
q^0_n &=&   q^0_{n+1} - \frac{C_{n+1}}{\alpha
\p{p}{q}{1}{n+1}},
\\
q^1_n &=&  q^1_{n+1} - \frac{C_{n+1}}{\alpha},
\\
\p{p}{q}{0}{n} &=& \p{p}{q}{0}{n+1},
\\
\p{p}{q}{1}{n}&=& \p{p}{q}{1}{n+1} 
+ \frac{C_{n+1}}{\p{p}{q}{1}{n+1}},
\end{eqnarray}
and the Lagrange multipliers get determined to be,
\begin{equation}
  N_n = \frac{C_{n+1}}{\alpha \p{p}{q}{1}{n+1}},
\end{equation}
where $C_{n+1}=\p{p}{q}{0}{n+1} + (\p{p}{q}{1}{n+1})^2/2 +\alpha
q^1_{n+1}$. The evolution scheme runs backwards, one can construct a
scheme that runs forward by solving for $N$ and $\pi$ at instant $n$
when imposing the constraints strongly.
The two
methods yield evolution schemes of different functional form
since one propagates ``forward'' in time and the other ``backward''.
The inequivalence in the functional form stems from the fact that
the discretization of the time derivatives chosen in the Lagrangian
is not centered. It should be emphasized that if one starts from
given initial data and propagates forwards with the first system of
equations and then backwards using the second, one will return to
the same initial data.

So we see in the example how the mechanism works. It yields evolution
equations that usually are implicit as evolution schemes. The
equations are consistent. The Lagrange multipliers get determined by
the scheme and there are no constraints left on the canonical
variables. The evolution is implemented by a (non-singular) canonical
transformation. The number of degrees of freedom is larger than those
in the continuum. There will exist different sets of initial data that
lead to different solutions for the discrete theory but nevertheless
will just correspond to different discrete approximations and
parameterizations of a single evolution of the continuum theory.

The example also exhibits some of the problems the framework may face.
Since one is generically left with implicit non-linear systems of
algebraic equations to be solved, there is no guarantee that the
solutions will be real. Neither is it guaranteed that the Lagrange
multipliers will be bounded in value (this is important since in
totally constrained theories the lapse controls the evolution step and
one wishes it to be small in the continuum limit). Although we
explored some simple examples where this does not pose a problem in
several publications, it is clear that generically it could be a
serious obstacle. This obstacle can be related to the intuitive fear
that people manifested about the formalism being a ``gauge fixing''.
Gauge fixed quantizations are not favored since it is known that one
cannot fix gauges globally. Similarly here, the framework cannot
generically guarantee that one approximates the continuum globally in
a solution. We will now argue that some simple, yet carefully chosen,
discretizations actually achieve this goal. This is in principle 
achieved in a generic fashion, i.e. we can find such discretizations
for any given totally constrained system.

\subsection{Uniform discretizations: the key idea}

As we stressed before, there is significant freedom in how one chooses
to discretize the action. We will now use this freedom to our
advantage, as was first proposed in \cite{uniform}. Consider a totally
constrained system with $N$ first class constraints $\phi_i$ and a
configuration space with $2P$ variables $(q,p)$ (we omit indices on
the variables for simplicity). We will assume that one has chosen the
discretization of the action in such a way that the evolution
equations for a given dynamical variable $A$ of the theory can be
written as,
\begin{equation}
A_{n+1} = e^{\{\bullet,H\}}(A_n)\equiv A_n +\{A_n,H\}
+{1 \over 2}\{\{A_n,H\},H\}+\cdots
\end{equation}
As is obvious, $H$ is a constant of the motion of the discrete evolution.
The quantity $H$ is defined in the following way:
Consider a smooth function of $N$
variables $f(x_1,\ldots,x_N)$ such that the following three conditions
are satisfied: a) $f(x_1,\ldots,x_N)=0 \iff x_i=0 \forall i$ and
otherwise $f>0$; b) ${\partial{f}\over\partial{x_i}}(0,\ldots,0)=0$;
c) ${\rm det} {\partial^2 f \over \partial x_i \partial x_j}\neq 0$
$\forall x$ and d) $f(\phi_1(q,p),\ldots
\phi_N(q,p))$ is defined for all $(q,p)$ in the complete phase space. 
Given this we define $H(q,p)\equiv f(\phi_1(q,p),\ldots
\phi_N(q,p))$.

A particularly simple example is $H(q,p) ={1/2}\sum_{i=1}^N \phi_i(q,p)^2$,
a choice that has interesting parallels with the ``master constraint''
of the ``Phoenix project'' \cite{phoenix} as we shall discuss later.

An important point is that if we choose initial data such that
$H<\epsilon$ then $\phi_i$ remain bounded throughout the evolution and
will tend to zero in the limit $\epsilon\to 0$. Let us see that in
this limit one recovers the evolution equations given by the total
Hamiltonian $H_T$ in the constrained continuum theory. Let $H$ as in
the simple example above and take its initial value to be
$H_0=\delta^2/2$. We define $\lambda_i=\phi_i/\delta$, and therefore 
$\sum_{i=1}^N \lambda_i^2 =1$. The evolution of the dynamical
variable $q$ is given by,
\begin{equation}\label{22}
q_{n+1}=q_n +\sum_{i=1}^N\{q_n,\phi_i\} \lambda_i\delta +O(\delta^2)
\end{equation} 
and if we define $\dot{q}\equiv \lim_{\delta\to
0}(q_{n+1}-q_n)/\delta$, where we have identified the ``time
evolution'' step with the initial data choice for $\delta$, one then
has,
\begin{equation}
\dot{q} =\sum_{i=1}^N\{q,\phi_i\}\lambda_i, 
\end{equation}
and similarly for other dynamical variables. The specific values of
the multipliers $\lambda_i$ depend on the initial values of the
constraints $\phi_i$. 
Notice that taking the continuum limit requires
that the Lagrange multipliers be determined as is usual in the
consistent discretization approach, but are well defined bounded real 
functions of phase space, bypassing an important objection to the
original approach.

Notice that the procedure deals only with first class
constraints. In case of a system with second class constraints, it
will be needed to use Dirac brackets and work with the equivalent
first class system.

An important comment is needed here about the difference in the case
of a mechanical system and a field theory. Although it is true that
field theories formulated on the lattice are mechanical systems,
generically the constraints of field theories formulated on the
lattice fail to be first class, even if they were so in the
continuum. If the constraints of the discrete theory are not first
class the above procedure fails to reproduce the equations of the
continuum. Here one has two choices. One would be to first take the
continuum limit spatially keeping time discrete. In that case it may
happen that the resulting constraints of the theory with continuum
space and discrete time are first class. Then the above proof goes
through. The alternative is to consider in the discrete evolution
equation (\ref{22}) the Dirac bracket of the discrete theory. Then one
can show that the continuum limit is correctly achieved, including in
the Hamiltonian both first and second class constraints.

One could recast the current proposal in terms of the original
approach to consistent discretizations. There one started from an
action and noted that the Lagrangian could be viewed as a the
generating function of a canonical transformation between instants
$n$ and $n+1$. To be concrete, let us 
analyze a system with $N$ Abelian constraints.
We introduce the Lagrangian
\begin{equation}
L(q_n,q_{n+1},\lambda_1,\ldots,\lambda_N)=
S(q_n,q_{n+1},\lambda_1,\ldots,\lambda_N)
+g(\lambda_1,\ldots,\lambda_N)
\end{equation}
where $L$ is a type 1 generating function of a canonical
transformation between canonical variables $q_n,p_n$ and
$q_{n+1},p_{n+1}$, $S$ is Hamilton's principal function for a given
set of Lagrange multipliers $\lambda_1,\ldots,\lambda_N$ (they are
evaluated at instant $n$, we omit the subscript for simplicity),
$g$ is such that $g(0)=0$ and the mappings $\lambda_i\to {\partial g \over
\partial \lambda_i}$ and $x_i \to {\partial f \over \partial x_i}$ are
inverse where $f$ is the function used to define the Hamiltonian.  The
generating function yields the canonical momenta in the usual way
$p_{n+1}=\partial L /\partial q_{n+1}$, $p_n=-\partial L /\partial
q_n$. One also has that $\partial L/\partial \lambda_i=0$ and this
determines the Lagrange multipliers, $\lambda_i=h_i(\phi)$, where
$h_i$ is the inverse function of the mapping defined by
$\lambda_i\rightarrow {{\partial{g}}\over{\partial{\lambda_i}}}$.
This evolution corresponds to a Hamiltonian
$H=f(\phi_1,\ldots,\phi_N)$, with $\partial_i{f}=h_i$.  In particular,
the simplest case is when $g=\sum_{i=1}^N x_i^2/2$ and then
$H=\sum_{i=1}^N \phi_i^2/2$.  The generating function $L$ allows to
determine the discrete evolution that preserves exactly the value of
the constraints of the continuum theory and recovers the continuum
limit when all $\phi_i\to 0$ in the initial data. We therefore see
that the approach proposed here is a particular case of the consistent
discretizations. We have just chosen to discretize things in a way
that the Hamiltonian is simple ---rather than the action--- and this
guarantees a good continuum limit.

The constants of the motion of the discrete theory are quantities that
have vanishing Poisson bracket with the Hamiltonian, $\{O^D_i,H\}=0$
and in the continuum limit $H_0 \to 0$ reproduce, as functions of
phase space the ``perennials'' of the continuum theory
$O^C_i=\lim_{H_0\to 0} O^D_i$. This can be immediately seen from the
fact that the discrete equations reproduce the continuum equations for
any dynamical variable in the continuum limit. Conversely, for every
perennial of the continuum theory there exists a constant of the
motion (in general many constants) of the discrete theory that reduce
to the given perennial in the continuum limit.
We have therefore shown that uniform discretizations recover the constraints
and the perennials of the continuum theory and therefore provide a good
starting point for a quantization of the continuum theory.

\subsection{Quantization}

We now turn our attention to the quantum theory. We will introduce a 
Heisenberg quantization for the discrete theory (this is more natural
given that one has an explicit evolution).
To quantize the theory we follow several steps. We start with the classical
discrete system constructed as in the previous section, we then have  the
canonical variables at level $n+1$ in terms of the variables at level $n$,
$
q_{n+1} = q_{n+1}(q_n,p_n), \quad p_{n+1} =p_{n+1}(q_n,p_n)$.

We then define the
kinematical space of states of the quantum theory, ${\cal H}_k$, as 
the space of functions of $N$ real variables $\psi(q)$ that are square
integrable. In this space we define operators $\hat{Q}$ and $\hat{P}$
as usual.
To construct the operators at other time levels (in the Heisenberg
Picture) we introduce a unitary operator $\hat{U}$ that we will
define later that gives,
\begin{equation}
\hat{Q}_{n} \equiv \hat{U}^{-1} \hat{Q}_{n-1} \hat{U} = 
\hat{U}^{-n} \hat{Q}_0
\, \hat{U}^n\,, \quad \hat{P}_{n} \equiv \hat{U}^{-1} 
\hat{P}_{n-1} \hat{U} =
\hat{U}^{-n} \hat{P}_0 \,\hat{U}^n\,.
\end{equation}

When the evolution is determined by a discrete Hamiltonian $H$, as
is the case in the uniform discretizations, the evolution operator is
given by $\hat{U}=e^{-i\hat{H}/\hbar}$. Notice that $\hat{U}$ may also
be determined by requiring that the fundamental operators satisfy an
operatorial version of the evolution equations,
\begin{equation}\label{Cuant-005}
\hat{Q}_{n}\, \hat{U} - \hat{U} Q_{n+1}(\hat{Q}_n,\hat{P}_n)=0, 
\quad \hat{P}_n\, 
\hat{U} - \hat{U}
P_{n+1}(\hat{Q}_n,\hat{P}_n)=0,
\end{equation}
and this provides a consistency criterion for the construction of $\hat{U}$.

At a classical level $H=0$ if and only if the constraints $\phi_i=0$.
There exists a natural definition of the physical space of the
continuum theory that does not require that we refer to the
constraint. Since we know that $\hat{U}=\exp(-i \hat{H}/\hbar)$, a necessary
condition satisfied by the states of the physical space of the
continuum theory, $\psi\in{\cal H}_{\rm phys}$ is given by $\hat{U}\psi =
\psi$. More precisely the states $\psi$ of  ${\cal H}_{\rm phys}$ should 
belong to the dual of a space $\Phi$ of functions sufficiently 
regular on ${\cal H}_k$.
That is, the states $\psi\in {\cal H}_{\rm phys}$ satisfy 
\begin{equation}
\int \psi^* \hat{U}^\dagger \varphi dq = \int \psi^* \varphi dq,
\end{equation}
where $\varphi\in \Phi$. 
This condition characterizes the quantum physical space 
of a constrained continuum theory without needing to implement the 
constraints as quantum operators by using the discretization technique.

The unitary operators of the discrete theory allow to
construct the ``projectors'' onto the physical space of the continuum
theory, which is one of the main goals of any quantization procedure
based on Dirac's ideas. It should be noted that these are really 
generalized projectors in the sense that they project to a set of functions
that belong in the dual of a subspace of sufficiently well behaved
functions of ${\cal H}_k$. All of this is achieved without having to
define the quantum constraint. To construct the ``projectors'' one can
compute,
\begin{equation}\label{projectorcontinuum}
\hat{P}\equiv \lim_{M\to\infty} C_M \hat{U}^M.
\end{equation}
If such a limit exists for some $C_M$ such that 
$\lim_{M\to\infty}(C_{M+1}/C_M)=1$ then $\hat{U} \hat{P} = \hat{P}$, and 
we have that
$\hat{U} \hat{P} \psi = \hat{P}
\psi$,  $\forall \psi \in {\cal H}_{\rm k}$.

The limit exists in several examples in which $\hat{H}$ has a continuum
spectrum, as we shall see. If the spectrum is discrete with
eigenvalues $e_i$ and it contains a vanishing eigenvalue $e_{i_0}$
then a projector is trivially defined as $|e_{i_0}>< e_{i_0}|$. A
constructive procedure leading to a general definition of the
projector in terms of the discrete evolution operator $\hat{U}$ valid for
any spectrum, continuum or discrete, is given by:
\begin{equation}\label{projector}
\hat{P}\equiv \lim_{M\to\infty} \sum_{n=M}^{{\rm Int}(rM)}
{{C_n \hat{U}^n}\over{{\rm Int}(rM)-M}}.
\end{equation}
where $r$ is a real number grater than one and ${\rm Int}(rM)$ is the
integer part of $rM$. If $\hat{U}$ has a continuum spectrum this
definition is a trivial consequence of the previous one. In the case
of a discrete spectrum one can check that the definition works
recalling the definition of the Kronecker delta in terms of a Fourier
series.  Notice that the definition of physical space that Thiemann
introduces in the ``Phoenix project'' \cite{phoenix}, is equivalent to
the choice we make if one is considering the Hamiltonian that is
quadratic in the constraints.  Furthermore, given two states of ${\cal
  H}_{\rm phys}$, $\psi_{\rm ph}$, $\phi_{\rm ph}$, where $\psi_{\rm
  ph}=\hat{P} \psi$, and $\phi_{\rm ph}=\hat{P} \phi$, the physical
inner product is defined by $<\psi_{\rm ph}|\phi_{\rm ph}>= \int dq
\phi(q)^* \hat{P} \psi(q)$ and a physical inner product is determined
by the projector constructed from the discrete theory.

We now illustrate the technique with a rather general example. We consider
a generic mechanical system with a finite dimensional phase space
with one constraint $\phi=0$. We will show that the projector 
constructed with our technique reproduces the one constructed with
group averaging techniques \cite{groupaveraging}. That is,
\begin{equation}
P =\lim_{M\to \infty} \sqrt{i M \over \pi} 
e^{-i M \phi^2} = \int_{-\infty}^\infty
{d\mu \over 2 \pi} e^{i \mu \phi}.  
\end{equation}
To make contact with the group averaging case we need to 
assume that $\phi$ is a self-adjoint operator in the kinematical
phase space with an eigenbasis given by 
$\phi |\alpha> = \phi(\alpha) |\alpha>$ and $1=\int |\alpha><\alpha| d\alpha$.
The proof of the equivalence is,
\begin{equation}
P =P \int |\alpha><\alpha| d\alpha = \int \lim_{M\to\infty} \sqrt{i M \over \pi}
e^{-i M \phi^2} |\alpha><\alpha|
\end{equation}
and noting that $\lim_{M\to \infty} \sqrt{i M\over \pi} 
e^{-i M x^2} =\delta(x) = 
\int_{-\infty}^{\infty} {d \mu \over 2 \pi}e^{i\mu x}$ the proof is complete.
For the proof we assumed a quadratic form of the Hamiltonian, but can
actually be extended to Hamiltonians of the general form we discussed above,
computing the integral by steepest descents. The proof can also be extended
to systems with $N$ Abelian constraints by noting that $\lim_{M\to \infty} 
({i M\over \pi})^{N/2} e^{-i M {\vec x}.{\vec x}} =\delta(\vec x) = 
\int_{-\infty}^{\infty} {d \mu^N }e^{i{\vec \mu}.{\vec x}}/(2 \pi)^N$. 

A point to be noted is that if the Hamiltonian is such that zero is not
an eigenvalue, the above expression for the projector vanishes. Therefore
the technique cannot be applied. We will confront a situation like this
in section VI. There we will argue that in such models one should take
a different point of view than defining the physical Hilbert space in
the traditional way.

\section{Example 1: Homogeneous isotropic cosmology}

We will follow the same procedure as in ordinary loop quantum
cosmology (see \cite{lqc} for a review).  We start by considering
general relativity formulated in terms of Ashtekar's variables. We
then assume that we have a homogeneous and isotropic universe with
flat spatial sections. In such case one can choose the triad and
Ashtekar connection as,
\begin{equation}
    A^i_a(x) = {c \delta^i_a \over V_0^{1/3}},\qquad E^a_i = {p \delta^a_i \over V_0^{2/3}}.
\end{equation}
with $\{c,p\} = (8\pi/3) G$ where we have taken the Immirzi
parameter as unity. As usual one proceeds in analogy with the full
theory and defines the holonomy along an ``edge'' of ``length''
$\mu V_0^{1/3}$, where $V_0$ is the volume an elementary cell of
universe,
\begin{equation}
    h^{(\mu)}_i(c)= \cos(\mu c/2) +2 \tau_i \sin(\mu c/2).
\end{equation}

The loop quantum cosmology approach, roughly speaking, can be characterized
as attempting to produce for the homogeneous context expressions that
mimic those that are found in the loop representation for the full theory.
In particular, one takes the usual Hamiltonian constraint of the
classical theory and represents the quantities involving connections
using holonomies and the expressions involving the triads using the
Poisson bracket of the volume with the connection, as first noted by
Thiemann in the full theory \cite{qsd}.

The resulting expression for the Hamiltonian is,
written as
\begin{equation}
  C_{\rm grav}^{(\mu_{0})} =-{4 {\rm sgn}(p)\over 8\pi \mu_0^3 G}
\sum_{ijk} \epsilon^{ijk}{\rm Tr}\left(
h_i^{(\mu_0)}h_j^{(\mu_0)}h_i^{(\mu_0)-1}h_k^{(\mu_0)}\left\{
h_k^{(\mu_0)-1},V\right\}\right)+ O(c^3 \mu_0),
\end{equation}
where $V=|p|^{3/2}$. In the continuum limit $(\mu_0\to 0)$ it can be shown
that the expression reproduces the Hamiltonian constraint of a
homogeneous isotropic vacuum minisuperspace.

It turns out to be more convenient to evaluate explicitly the Poisson
bracket and replace the holonomies by the expression above to get,
\begin{equation}
 C_{\rm grav}^{(\mu_{0})} =-\sqrt{6} {\sqrt{16\pi G |p|}\over\mu_0^2}
\sin^2\left({\mu_0 c}\right).
\end{equation}
Where we have made the redefinition ${3 \over 8 \pi G}p \rightarrow
p$ so that $\{c,p\} = 1$, we will use this redefinition of $p$ in
what follows. This Hamiltonian constraint can be obtained from a
Lagrangian for gravity coupled to a scalar field of the form,
\begin{equation}
  L = p\, \dot{c} + p_\phi \dot{\phi} + {M\over 2} 
\left[{\sqrt{|p|}\sin^2(\mu_0 c) \over \mu_0^2}
- {{p_\phi^2 } \over |p|^{3/2}} \right].
\end{equation}
$M$ is the Lagrange multiplier (proportional to the lapse) with
the factor $\sqrt{6}\sqrt{16 \pi G}$ absorbed in it. For
simplicity we have also rescaled the scalar field and its
conjugate so the gravitational coupling $3/(16\pi G)$ is absorbed.

We immediately recognize in the classical limit $\mu_0\to 0$ the
usual Lagrangian for the Friedmann--Robertson--Walker model
coupled to a scalar field.

Given that we are dealing with a mechanical system with one
constraint, one can immediately apply the technique used in the 
example of the last section to show that one recovers the same
physical space as in the group averaging approach. We therefore
make contact exactly with traditional loop quantum cosmology, 
at the level of physical space.

It would be interesting to pursue this example further by not
taking the continuum limit. There one could introduce a relational
time and study loss of unitarity and other effects that will arise,
but we will postpone a detailed discussion of cosmology 
to a subsequent paper.

\section{Example 2: A model with Abelian constraints}

We consider a mechanical system 
 with configuration manifold $R^N$, 
coordinatized by $q^i,i=1,\ldots,N$ and $M<N$ commuting constraints
\begin{equation}
C^i=p^i \qquad i=1,\ldots,M \quad,
\end{equation}
where the $p^i$'s are the conjugated momenta to the $q^i$'s.

All phase space functions which do not depend on the first $M$ $q^i$'s 
are Dirac
observables, i.e. functions which commute with the constraints 
(on the constraint hypersurface). 

A Dirac observable which depends on
$p^i,\,i=1,\ldots,M$ is equivalent to the Dirac observable obtained
from the first one by setting $p^i=0$ (since these two observables
will coincide on the constraint hypersurface). Therefore it is
sufficient to consider observables which are independent of the first
$M$ configuration observables and of the first $M$ conjugate momenta.
A canonical choice for an observable algebra is the one generated by
$q^i,p^i,\,i=(M+1),\ldots,N$.

Let us now consider the uniform discretization. For that we construct
$H ={k \over 2} \sum_{i=1}^M {(p^i)^2}$. The constant $k$ is to make the
Hamiltonian a dynamical variable with units of action. The discrete
evolution equations are,
\begin{eqnarray}
  q^a_{n+1} &=& q^a_n + k \sum_{b=1}^M \left\{ q^a_n,p^b_n\right\} p^b_n,\\
  &=& q^a_n + k p^a_n \Theta_{a,M},\;\text{(there is no summation in $a$)}\\
  p^a_{n+1}&=&p^a_n,
\end{eqnarray}
where $\Theta_{a,M}=1$ if $a \le M$ and zero otherwise.

To take the classical continuum limit, we assume that $H=\delta^2/2$ with
$\delta\to 0$. If we define 
\begin{equation}
\dot{q}^a\equiv \lim_{\delta \to 0} {q^a_{n+1}-q_{n}\over \delta}
\end{equation}
we have from the evolution equations that,
\begin{eqnarray}
  \dot{q}^a &=&\lambda^a \Theta_{aM},\\
  \dot{p}^a &=&0,
\end{eqnarray}
where $\lambda^a\equiv \lim_{\delta \to 0} k p^a/\delta$.

To quantize the system, we will start with an auxiliary Hilbert space
$L_2(R^N)$ on which the operators $\hat q^i$ act as multiplication
operators and the momenta as derivatives,
\begin{eqnarray} 
\hat q^i \,\psi(q)&=&q^i
\,\psi(q)\\
\hat p^i\psi(q)&=&-i\hbar \,\partial_i \psi(q),
\end{eqnarray}
and construct a unitary operator,
\begin{equation}
  \hat{U} = \exp\left(-i {k\over 2\hbar} \sum_{i=1}^M \hat{p}_i^2\right),
\end{equation}
in terms of which one has
\begin{eqnarray}
  \hat{q}^a_{n+1} &=& U^\dagger \hat{q}^a_n U\\
&=& \hat{q}^a_n +\sum_{b=1}^M {k \over 2 \hbar} \left[(p^b_n)^2,q^a_n\right]\\
                  &=&\hat{q}^a_n+ k \hat{p}^a_n \Theta_{aM},
\end{eqnarray}
so we see that we have recovered the classical equations as 
operatorial relations in the Heisenberg representation. 

We now wish to define the physical space of states of the continuum
theory. To do this we will use the projector technique,
\begin{equation}
\hat{P}\equiv \lim_{R\to\infty} C_R \hat{U}^R.
\end{equation}
with 
$C_R=\left( \frac{i R k}{2 \pi \hbar} \right)^{M/2}$.
Acting on the identity written as 
$I=\int dp^1\cdots dp^N |p^1,\ldots,p^N><p^1,\ldots,p^N|$
we have,
\begin{eqnarray}
  \hat{P} &=&\hat{P}I= \int dp^1\cdots dp^N \lim_{R\to\infty} C_R
\exp\left(-{i R k\over 2\hbar}\sum_{a=1}^M
(p^a)^2\right)|p^1,\ldots,p^N><p^1,\ldots,p^N|\nonumber\\
&=& 
\int dp^{M+1}\cdots dp^N
|0,\ldots,p^{M+1},\ldots,p^N><0,\ldots,p^{M+1},\ldots,p^N|.
\end{eqnarray}

We therefore see that the physical space is given by the square
integrable functions depending on the variables $M+1$ to $N$. The
physical inner product is therefore immediately induced by the
kinematical inner product on the physical space of states. 

We would like now to turn our attention to a quantization where
we assume the continuum limit is not taken, i.e. we end up with 
a discrete quantum theory. Although this is not necessary in this
example, it would be in examples where the continuum limit is 
problematic. It is therefore good to see how the discrete quantum
theory can be viewed as a quantizations of the continuum theory,
at least at certain scales. Since we have a conserved Hamiltonian
if we choose an initial quantum state that has a small expectation
value for the Hamiltonian, it will keep its small expectation value
upon the unitary evolution defined by $\hat{U}$. We will take 
advantage of working in the discrete theory to be able to measure
variables that would not be observable in the continuum theory
and in terms of them define a relational evolution that will 
approximate well the dynamics of the continuum theory. 

We start by recalling one way in which the dynamics of models like
this can be handled in the continuum via the use of ``evolving
constants'' \cite{rovelli}. Given any function of the dynamical
variables of the theory $f(q^1,\ldots,q^N,p^1,\ldots,p^N)$ one can
define an ``evolving constant''. This is a function that depends on a
given number of real parameters
$f(T^1,\ldots,T^M,q^{M+1},\ldots,q^N,p^{M+1},\ldots,p^N)$ and that
coincides with the quantity of interest on the constraint surface when
the partial observables $q^1,\ldots,q^M$ take the values
$T^1,\ldots,T^M$. The ``evolving constant'' is a physical observable
of the theory, that is, it has vanishing Poisson brackets with the
constraints. To determine the dynamics of the theory is to determine
what is the value of the quantity of interest $f$ for a given value of
the parameters $T^1,\ldots,T^M$. A potential drawback of this approach
is that it requires to treat some of the variables of the problem as
classical variables, something that may not appear very natural in
certain regimes where all variables should exhibit quantum mechanical
behaviors.

We now go back to the discrete theory. Since the theory has no
constraints we can work in the kinematical space and all variables
will be quantum mechanical in nature. We wish to describe the
evolution relationally, defining conditional probabilities (see
for instance \cite{njp}).
The physical variable we will consider is
$\hat{F}\equiv \hat{q}^1\hat{p}^N$. This operator coincides with the
angular momentum $L_{1 N}=\hat{q}^1\hat{p}^N-\hat{q}^N\hat{p}^1$
evaluated on the constraint surface. 
We now need to choose a ``clock''. Here we need to be careful. If we 
choose $q^1,\ldots,q^M$ as ``clock'' variables the problem
that may arise is that if one makes a very precise measurement of 
one of these variables would imply a large change in the expectation
value of the Hamiltonian that would lead to lose the regime close
to the continuum. This is understandable, in a theory with fundamental
discreteness there has to be a limit as to how accurate a measurement
one can make. For instance, this would be akin to claiming that in 
a discrete theory of quantum gravity one cannot measure lengths shorter
than the Planck length. It should be recalled that this is also a rather
artificial limitation. If one is dealing with a realistic theory that
includes measuring devices in it, since the evolution is unitary, one
is guaranteed to remain in the regime one chooses irrespective of 
what measurements are made with the devices the theory incorporates.
One can only break the regime if one introduces measuring devices
that are external to the theory. In the particular example we are
considering the theory is too simple to encompass measuring devices
within it. We are therefore forced to consider external measurement
devices and therefore we need to limit their interactions with the
model to respect the choice of initial conditions within the regime
of being close to the continuum.

Taking the last point into account, 
the projector for a clock reading a value $(q^1,\ldots q^M)$, where
we understand that each variable takes a value within an
interval $\Delta q$ centered at each $q^a$, is
\begin{equation}
P_q = \int_\Delta dq^1\cdots dq^M \int dp^{M+1}\cdots dp^N |q^1,\ldots,q^{M},
p^{M+1},\ldots,p^N> <q^1,\ldots,q^{M}, p^{M+1},\ldots,p^N|
\end{equation}
and the projector for a measurement of $\hat{F}$ to be in an
interval $(f_1,f_2)$ is,
\begin{equation}
P_F = \int dq^1\cdots dq^M dp^{M+1}\cdots dp^N |q^1,\ldots,q^{M},
p^{M+1},\ldots,p^N> <q^1,\ldots,q^{M}, p^{M+1},\ldots,p^N|
\theta(f_1/q^1,p^N,f_2/q^1)
\end{equation}
where $\theta(a,b,c)=1$ if $a\leq b \leq c$ and zero otherwise.

The projector for a simultaneous measurement of the clock and
$\hat{F}$ therefore is,
\begin{equation}
P_q P_F = \int_\Delta dq^1\cdots dq^M \int dp^{M+1}\cdots dp^N |q^1,\ldots,q^{M},
p^{M+1},\ldots,p^N> <q^1,\ldots,q^{M}, p^{M+1},\ldots,p^N|
\theta(f_1/q^1,p^N,f_2/q^1).
\end{equation}

Let us consider a state at instant $n=0$ of the form,
\begin{equation}
| \psi_0> = \int dp^{1}\cdots dp^N
c_{p^1,\ldots,p^N}|p^1,\ldots,p^N>
\end{equation}
therefore, upon evolution,
\begin{equation}
| \psi_n> = \hat{U}^n  | \psi_0> = \int dp^{1}\cdots dp^N
c_{p^1,\ldots,p^N} \; \text{exp}\left( -\frac{ikn}{2\hbar}
\sum_{i=1}^M  (p^i)^2\right) |p^1,\ldots,p^N>.
\end{equation}

The conditional probability is given by
$\mathcal{P}(F|q)=
(\sum_n < \psi_n| P_q P_F | \psi_n>)/(\sum_n < \psi_n| P_q |
\psi_n>)$ that is, the ratio of the probability 
that the clock measures $q$ and the system
under study takes the value $F$ simultaneously and the probability
that the clock measures $q$ \cite{njp}. Explicitly,
\begin{equation}
\mathcal{P}(F|q) = \frac{\sum_n \int_\Delta dq^1\cdots dq^M 
\int dp^{M+1}\cdots dp^N
\theta(f_1/q^1,p^N,f_2/q^1) |D_n(q^1,\ldots,q^{M},
p^{M+1},\ldots,p^N)|^2}{\sum_n \int_\Delta dq^1\cdots dq^M 
\int dp^{M+1}\cdots dp^N
 |D_n(q^1,\ldots,q^{M},
p^{M+1},\ldots,p^N)|^2 }
\end{equation}
where we have defined
\begin{equation}
D_n(q^1,\ldots,q^{M}, p^{M+1},\ldots,p^N) \equiv \int dp^{1}\cdots
dp^M c_{p^1,\ldots,p^N} \; \prod_{j=1}^M <q^j|p^j>
\text{exp}\left( -\frac{ikn}{2\hbar} (p^j)^2\right)\,.
\end{equation}

If we assume that the initial state at $n=0$ takes a tensorial
product form (which means the ``clock'' is not interacting with
the ``system under study'') $c_{p^1,\ldots,p^N} =
a_{p^1,\ldots,p^M} b_{p^{M+1},\ldots,p^N}$ we have
\begin{equation}\label{prodtensorial}
\frac{ \int_\Delta dq^1\cdots dq^M \int dp^{M+1}\cdots dp^N
\theta(f_1/q^1,p^N,f_2/q^1) |b_{p^{M+1},\ldots,p^N}|^2
h(q^1,\ldots, q^M)}{ \int_\Delta dq^1\cdots dq^M \int
dp^{M+1}\cdots dp^N
 |b_{p^{M+1},\ldots,p^N}|^2
h(q^1,\ldots, q^M) }
\end{equation}
for the conditional probability. Here we have defined
\begin{equation}
h(q^1,\ldots,q^{M}) \equiv \sum_n \left| \int dp^{1}\cdots dp^M
a_{p^1,\ldots,p^M} \; \prod_{j=1}^M <q^j|p^j> \text{exp}\left(
-\frac{ikn}{2\hbar} (p^j)^2\right) \right|^2\,.
\end{equation}

Since we are interested in comparing the results of our approach
with that of the evolving constant approach in the continuum, we will
now take the continuum limit. It is given by,
$a_{p^1,\ldots,p^M}=\delta(p^1)\cdots\delta(p^M)$ the
$h(q^1,\ldots, q^M)$ become constants and the probability is
\begin{equation}\label{probcontinuo}
\mathcal{P}(F|q) = \frac{\int_\Delta  dq^1 \int dp^{M+1}\cdots
dp^N \theta(f_1/q^1,p^N,f_2/q^1) |b(p^{M+1},\ldots,p^N)|^2}{\Delta
\int dp^{M+1}\cdots dp^N
 |b(p^{M+1},\ldots,p^N)|^2}
\end{equation}
This completes the calculation in the relational picture.

Let us now revert, for comparison, to the
``evolving constant'' picture. To do this we need to construct
a state of the physical space of the continuum theory such that
it yields the same expectation value for the observables
that $|\psi_0>$. It is given by a density matrix,
\begin{equation}
\rho_{\rm phys}  = {\rm Tr}|_{q^1\ldots q^M} |\psi_0><\psi_0|.
\end{equation}
The physical quantity to measure
is $\hat{F}\equiv q^1 \hat{p}^N$ (where $q^1$ is viewed as a c-number and
therefore this quantity is an observable). The probability of getting a value
in the interval $(f_1,f_2)$ is
\begin{eqnarray}
\mathcal{P}& \equiv& \frac{{\rm Tr}\left(\int_\Delta dq^1
\int dp^1\cdots dp^N
|p^1,\ldots,p^N> <p^1,\ldots,p^N|
\theta(f_1/q^1,p^N,f_2/q^1)\rho_{\rm phys}\right)}
{\Delta {\rm Tr}\left(\rho_{\rm phys}\right)}\\
& = &
\frac{\int_\Delta  dq^1 \int dp^{M+1}\cdots dp^N
\theta(f_1/q^1,p^N,f_2/q^1) |b(p^{M+1},\ldots,p^N)|^2}{\Delta \int
dp^{M+1}\cdots dp^N
 |b(p^{M+1},\ldots,p^N)|^2}
\end{eqnarray}
and in the case in which the state is a tensor product of the form
$c_{p^1,\ldots,p^N} = a_{p^1,\ldots,p^M} b_{p^{M+1},\ldots,p^N}$
this probability indeed coincides with (\ref{probcontinuo}). 

We have therefore seen how one can work in the discrete theory to
introduce a relational notion of time in the quantization of the
discrete theory and how the physical predictions of this picture
coincide with those of the ``evolving constants'' picture in the
continuum theory if one takes the continuum limit. The agreement is
qualified in the sense that we cannot carry out measurements of
arbitrary precision of ``classical'' parameter of the evolving
constant. This is to be expected in a treatment where all variables
are treated quantum mechanically in an equal footing without singling
out one to behave classically.

\section{Example 3: A system with compact gauge group and continuum limit}

Here we will consider a system with non-Abelian constraints that 
form a Lie algebra. The model is given by the same phase space
as in the previous section restricted to three configuration space
dimensions.  We consider a mechanical system 
with configuration manifold $R^3$, 
coordinatized by $q^i,i=1,2,3$ and 3 non-commuting constraints
\begin{equation}
C^i=L^i\equiv \epsilon^{ijk} q^j p^k \qquad i=1,2,3 \quad,
\end{equation}
where the $p^i$'s are the momenta conjugate to the $q^i$'s and 
we assume Einstein's summation convention on repeated indices. Notice
that the three constraints are not independent. Vanishing angular 
momentum is equivalent to requiring $q^i=\lambda p^i$ with $\lambda$
an arbitrary number, which implies two conditions between the 
phase space variables. The constraint surface is four dimensional
and the system will have two independent observables. In our approach
we could choose to construct the Hamiltonian starting from two
independent constraints or one could choose to use a more symmetric,
yet redundant, form of the constraints. We will choose the latter.
We introduce a canonical evolution through a Hamiltonian,
\begin{equation}
  H=\sum_{i=1}^3 {(L^i)^2\over 2 k},
\end{equation}
where $k$ is a constant with units of action. The discrete
evolution equations are,
\begin{eqnarray}
  q^i_{n+1} &=& q^i_n +\epsilon_{ijk}{ q^k L^j\over k} 
+\epsilon_{ijk}\epsilon_{jmn} q^m {L^n L^k \over k^2} +\ldots\\
  p^i_{n+1} &=& p^i_n +\epsilon_{ijk} {p^k L^j \over k}
+\epsilon_{ijk}\epsilon_{jmn} p^m {L^n L^k \over k^2}+\ldots
\end{eqnarray}

The continuum limit is obtained setting $H=\delta^2/2$ and defining
$\lambda^i = L^i/(\delta\sqrt{k})$ we have, taking the limit as
explained before,
\begin{eqnarray}
  \dot{q}^i &=& \epsilon_{ijk} {q^k \lambda^j \over \sqrt{k}},\\
  \dot{p}^i &=& \epsilon_{ijk} {p^k \lambda^j \over \sqrt{k}}.
\end{eqnarray}

We note that the components $L^i$ are constants of the motion and
therefore $\lambda^i$ are constant. There exist three further 
constants $q\cdot q=q^iq^i$, $p\cdot p=p^ip^i$, $q\cdot p = q^ip^i$. These are
not independent since $L^2-((q\cdot q)(p\cdot p)-(q\cdot p)^2)=0$. 
One therefore has
five independent constants of the motion of the discrete theory.
In the continuum limit the $L^i$'s vanish and one is left with
two independent constants of motion, for instance $q\cdot q$ and $q\cdot p$.
In the continuum theory, the trajectories are arbitrary trajectories on
a sphere. In the discrete theory, when one takes the continuum limit
one obtains trajectories that correspond to arbitrary circumferences
on the sphere, since the $\lambda^i$'s are constant. The constraint 
surface is therefore completely covered, but not all orbits of the
continuum theory. This however, does not cause problems since we
can recover all physical information of relevance with the trajectories
we get.

To quantize the model, we construct the same kinematical space as in
the previous example (restricted to a three dimensional configuration
space), and immediately construct the unitary operator,
\begin{equation}
  \hat{U}=\exp\left(-i {\hat{L}^2 \over 2 k \hbar}\right),
\end{equation}
which recovers as operatorial relations the classical discrete 
evolution equations (up to terms of order $\hbar$) so the
correspondence principle is satisfied.

To compute the projector we use the formula (\ref{projector}) and,
in terms of the a basis labeled by  the radial and angular momentum
eigenvalues,
\begin{equation}
  \hat{P} |n,\ell,m> = \delta_{\ell\,0} |n,\ell,m> =\delta_{\ell\,0} |n,0,0>,
\end{equation}
which can be rewritten as,
\begin{equation}
  \hat{P} = \sum_{n=0}^\infty |n,0,0><0,0,n|.
\end{equation}

The continuum limit is therefore trivially achieved. The physical
space is the space of zero angular momentum and there the constraints
of the continuum theory are automatically satisfied. We therefore
recover the traditional Dirac quantization.  The physical space is
given by the square integrable functions depending on the radial
variable $|q|$. The physical inner product is therefore immediately
induced by the kinematical inner product on the physical space of
states.

We will not work out in detail the quantization in which the
continuum limit is not taken, since it runs along similar lines
as what we discussed in the previous section. The only thing to
notice is that the current model would have a ``time variable''
consisting of two angles in the sphere. Again, one has 
restrictions on the accuracy with which one can measure angles 
so the approximation of the continuum theory is not lost.

\section{Example 4: A system with non-compact gauge group}

Here we will consider a system with non-Abelian constraints that
form a Lie algebra that is non-compact. More specifically, the
model is associated with the $SO(2,1)$ group. It has the same
phase space as in the previous section, except that the metric
will be given by ${\rm diag}(-1,1,1)$. The configuration variables
are $q^i$ with $i=0,1,2$ and the conjugate momenta are $p_i$. The
constraints are
\begin{equation}
C_i=L_i\equiv \epsilon_{ij}{}^k q^j p_k \qquad i=0,1,2 \quad,
\end{equation}
we assume Einstein's summation convention on repeated indices.
Here one has to be careful about upper and lower indices since the
metric is non-trivial. As in the previous example  we notice that
the three constraints are not independent. The constraint surface
is four dimensional and the system will have two independent
observables. In our approach we could choose to construct the
Hamiltonian starting from the two independent constraints or one
could choose to use a more symmetric, yet redundant, form of the
constraints. We will choose the latter.  We
introduce a canonical evolution through a Hamiltonian,
\begin{equation}
  H= {L_i L^i+2 L_0^2\over 2 k},
\end{equation}
where $L$ is a constant with units of action. The discrete
evolution equations are,
\begin{eqnarray}
  q^i_{n+1} &=& q^i_n +\epsilon^i{}_{jk} 
{q^k L^j\over k}-2\epsilon^i{}_{0k} {q^k L^0\over k} +\ldots\\
  p^i_{n+1} &=& p^i_n +\epsilon^i{}_{jk} 
{p^k L^j\over k}-2\epsilon^i{}_{0k} {p^k L^0\over k}+\ldots
\end{eqnarray}

The continuum limit is obtained setting $H=\delta^2/2$. Defining
$\lambda^i = L^i/(\delta\sqrt{k})$ for $i=1,2$ and $\lambda^0
=-L^0/(\delta\sqrt{k})$ we have, taking the limit as explained
before,
\begin{eqnarray}
  \dot{q}^i &=& \epsilon^i{}_{jk} {q^k \lambda^j \over \sqrt{k}},\\
  \dot{p}^i &=& \epsilon^i{}_{jk} {p^k \lambda^j \over \sqrt{k}}.
\end{eqnarray}

We note that although $L_0$ is a constant of the motion, the other
components are not. However $L_1^2+L_2^2$ is a constant, so $L_1$
and $L_2$ rotate around $L_0$ throughout the evolution.  There
exist three further constants $q\cdot q=q^i q_i$, $p\cdot
p=p^ip_i$ $q\cdot p = q^ip_i$. These are not independent since
$L\cdot L-((q\cdot q)(p\cdot p)-(q\cdot p)^2)=0$.  One therefore has
four independent constants of the motion of the discrete theory.
In the continuum limit two of these vanish and one is left with
two constants of motion, for instance $q\cdot q$ and $q\cdot p$.
In the continuum theory , the trajectories are arbitrary
trajectories on a two hyperboloids, one space-like and one
time-like. In the discrete theory, when one takes the continuum
limit one obtains trajectories that correspond to particular
choices of the Lagrange multipliers, depending on the initial
conditions chosen.

The quantization of this model (and similar ones) is known to be
problematic \cite{gomberoffmarolf, thiemann3, loukorovelli}. The core
of the problem is that if one promotes the constraints to operators on
the usual Hilbert space, they do not have a vanishing eigenvalue. This
can happen, but usually the resolution is to extend the Hilbert space
by including an improper basis of eigenvectors. This can be done in
this case, but one finds that the spectrum again does not contain
zero. More specifically, the continuum spectrum has eigenvalues larger
than $\hbar^2/4$.  One can find eigenvectors with eigenvalues smaller
than $\hbar^2/4$, but they do not arise as limit of functions of the
Hilbert space, i.e.  they cannot form part of an improper basis of the
Hilbert space. One can adopt the point of view that nevertheless the
eigenvectors with zero eigenvalue are the ``physical space'' of the
theory, essentially abandoning the idea that the physical space arises
as a suitable limit of the kinematical space. This was advocated by
\cite{gomberoffmarolf,loukorovelli}.  From the point of view of our
technique this is not satisfactory, since we wish to construct the
physical space of states starting from the quantum kinematical space,
taking a limit. As we mentioned before, lacking the zero eigenvalue in
the spectrum of the Hamiltonian yields our prescription for the
projector on the physical space useless.  With this in mind, the most
satisfactory solution is the one chosen in \cite{thiemann3}, where one
chooses as physical space the eigenvectors that have $\hbar^2/4$ as
eigenvalue. Another attractive possibility in the discrete approach is
{\em not} to take the continuum limit and retain a level of
fundamental discreteness. This is very natural in a model where it is
difficult to achieve a vanishing eigenvalue for the constraints, a
natural minimum existing for their eigenvalues. As we argued before,
such models can approximate the semiclassical physics of the theory of
interest with some restrictions on the type of states considered.
There can therefore be viewed as the best thing one can do in terms of
having an underlying quantum theory for the model that approximates
the classical physics of interest.

\section{Gravity in three dimensions}

In this section we will follow closely the construction of Noui and
Perez \cite{nouiperez}. We will therefore not repeat all the details
here. The difference will be that they construct the physical space
of states using group averaging and we will use the projector using
the technique we discussed earlier in this paper. 

We start with three dimensional gravity in first order { formalism}. {
  The space-time $\cal M$ is a three dimensional oriented smooth
  manifold} and the action is simply given by \begin{equation}
S[e,\omega]=\int_{\cal M} {\rm Tr}[{e} \wedge F({\omega})] \end{equation}
where
$e$ is the triad, i.e. a Lie algebra  valued $1$-form,
$F(\omega)$ is the curvature of the three dimensional connection
$\omega$ and $Tr$ denotes a Killing form on $\mathfrak g$. For
simplicity we will concentrate on Riemannian gravity so the previous
fields should be thought as defined on $SU(2)$ principal bundle over
${\cal M}$.

The phase space is parameterized by the pull back to $\Sigma$ of
$\omega$ and $e$. In local coordinates we can express them in terms of
the 2-dimensional connection $A_a^{i}$ and the triad field
$E^b_j=\epsilon^{bc} e^k_c \eta_{jk}$ where $a=1,2$ are space
coordinate indices and $i,j=1,2,3$ are $su(2)$ indices. The symplectic
structure is defined by
\[\{A_a^{i}(x), E^b_j(y)\}=\delta_a^{\, b} \; \delta^{i}_{\, j} \; \delta^{(2)}(x,y).\]
Local symmetries of the theory are generated by the first class constraints
$D_b E^b_j \simeq 0$ and $F_{ab}^i(A) \simeq 0$.

To quantize the theory we first find a
representation of the basic variables in an auxiliary Hilbert
space ${\cal H}_{\rm aux}$. The basic functionals of the connection are
represented by the set of holonomies along paths $\gamma \subset
\Sigma$. Given a connection $A$ and a path $\gamma$, one defines the
holonomy $h_{\gamma}[A]$ by
\begin{equation}
\label{hol}h_{\gamma}[A]=P \exp\int_{\gamma} A \;.
\end{equation}
 As for the triad, its associated basic variable is
given by the smearing of $E$ along co-dimension 1 surfaces. One
promotes these basic variables to operators acting on an auxiliary
Hilbert space where constraints are represented. The physical
Hilbert space is defined by those `states' that are annihilated by
the constraints. As these `states' are not normalizable with
respect to the auxiliary inner product they are not in 
${\cal H}_{\rm aux}$ and
have to be regarded rather as distributional.

The auxiliary Hilbert space is defined by the Cauchy completion of
the space of cylindrical functionals ${Cyl}$, on the space of
(generalized) connections $\bar {\cal A}$. The space $Cyl$ is defined as
follows: any element of $Cyl$, $\Psi_{\Gamma,f}[A]$ is a functional
of $A$ labeled by a finite graph $\Gamma \subset \Sigma$ and a
continuous function $f: SU(2)^{N_\ell({\Gamma})}\rightarrow C$
where $N_\ell({\Gamma})$ is the number of links of the graph
$\Gamma$. Such a functional is defined as follows
\begin{equation} \label{cyl}
\Psi_{\Gamma,f}[A]=f(h_{\gamma_1}[A],\cdots,h_{\gamma_{N_{\ell}(\Gamma)}}[A])
\end{equation}
where $h_{\gamma_i}[A]$ is the holonomy along the link $\gamma_i$ of
the graph $\Gamma$. { If one considers a new graph $\Gamma'$} such
that $\Gamma \subset \Gamma^{\prime}$, then any cylindrical function
$\Psi_{\Gamma,f}[A]$ trivially corresponds to a cylindrical function
$\Psi_{\Gamma^{\prime},f^{\prime}}[A]$. This space can be endowed with
the well known Ashtekar--Lewandowski inner product.  The Gauss
constraint can be defined in terms of the basic variables introduced
above. It generates gauge transformations whose action on $Cyl$
transforms the holonomy as follows
\begin{equation} \label{gg}
h_{\gamma}[A] \longmapsto  g_s h_{\gamma}[A]g_t^{-1}
\end{equation}
where $g_s,g_t\in SU(2)$ are group elements associated to the {\em
source} and {\em target} nodes of $\gamma$ respectively. The so-called
{\em kinematical} Hilbert space ${\cal H}_{\rm kin} \subset {\cal
H}_{\rm aux}$ is defined by the states in ${\cal H}_{\rm aux}$ which
are gauge invariant and hence in the kernel of the Gauss constraint.

A basis of gauge invariant functions can then be constructed by
contracting the tensor product of representation matrices with
$su(2)$-invariant tensors or $su(2)$-intertwiners. If we select an
orthonormal basis of intertwiners $\iota_n \in {\rm Inv}[j_1\otimes
j_2\otimes\cdots\otimes j_{
\scriptscriptstyle N_{\ell}}]$
where $n$ labels the elements of the basis we can write a basis of
gauge invariant elements of $Cyl$ called the {\em spin network
basis}. Each spin network is labeled by a graph $\Gamma \subset
\Sigma$, a set of spins $j_{\gamma}$ labeling links $\gamma$ of the
graph $\Gamma$ and a set of intertwiners $\iota_n$ labeling nodes $n$
of the graph $\Gamma$, namely:
\begin{equation}\label{spinnetwork}
s_{\scriptscriptstyle \Gamma,
\{j_{\ell}\},\{\iota_{n}\}}[A]=\bigotimes_{n \in
\Gamma} \ \iota_n\ 
\bigotimes_{\gamma \in \Gamma}\  
\stackrel{j_{\gamma}}{\Pi}(h_{\scriptscriptstyle
\gamma}[A]) \;.
\end{equation}
In order to lighten notations, we will omit indices (the graph,
representations and intertwiners) labeling spin-networks in the
sequel. One can take the kinematical inner product of two spin
networks $s,s'$, which we will denote $<s|s'>$ by integrating
expression (\ref{spinnetwork}) with the Ashtekar--Lewandowski measure.

Let us review the group averaging procedure to generate the projector,
following closely the discussion of Noui and Perez. One
starts with the formal expression
\begin{equation}
P =``\prod_{x \in \ \Sigma} \delta(\hat F(A))"=\int D[N] \ {\rm
exp}(i\int \limits_{\Sigma} {\rm Tr}[ N \hat {F}(A)])
\;,\label{ppp}
\end{equation}
where $N\in su(2)$.  We now introduce a regularization of
(\ref{ppp}). We will give a definition of $P$ by providing a
regularization of its matrix elements $<Ps,s^{\prime}>$ for any pair
of spin network states $s,s^{\prime} \in {\cal H}_{\rm kin}$. Let's
denote by $\Gamma$ and $\Gamma^{\prime}$ the graphs on which $s$ and
$s^{\prime}$ are defined respectively. We introduce an arbitrary
cellular decomposition of $\Sigma$ denoted $\Sigma^{\Gamma
\Gamma^{\prime}}_{\epsilon}$, where $\epsilon\in R$, such that
the graphs $\Gamma$ and $\Gamma^{\prime}$ are both contained in the
  graph defined by the union of $0$-cells and $1$-cells in $
{\Sigma^{\Gamma\Gamma^{\prime}}_{\epsilon}}$ and 
for each individual 2-cells (plaquette) $p$ there exist a
ball ${\cal B}_{\epsilon}$ of radius $\epsilon$---defined using
the local topology---such that $p\subset{\cal B}_{\epsilon}$.

Consequently all 2-cells shrink to zero when $\epsilon \rightarrow 0$.
We consider a local patch $U\subset \Sigma$ where we choose the
cellular decomposition to be square with cells of coordinate length
$\epsilon$. In that patch, the integral in the exponential in
(\ref{ppp}) can be written as a Riemann sum
\[F[N]=\int\limits_{U} {\rm Tr}[ N {F}(A)]=
\lim_{\epsilon\rightarrow 0}\ \sum_{p^i} \epsilon^2 {\rm
Tr}[N_{p^i} F_{p^i}],\] where $p^i$ labels the $i^{th}$ plaquette
and $N_{p^i}\in su(2)$ and $F_{p^i}\in su(2) $ are values of
$N^j\tau_j$ and $\tau_j \epsilon^{ab}F^{j}_{ab}[A]$ at some
interior point of the plaquette $p^i$ and $\epsilon^{ab}$ is the
Levi--Civita tensor. The basic observation is that the holonomy
$U_{p^i}\in SU(2)$ around the plaquette $p^i$ can be written as
\[U_{p^i}[A]={{\leavevmode\hbox{\small1\kern-3.8pt\normalsize1}}}
+ \epsilon^2 F_{p^i}(A)+{\cal O}(\epsilon^2)\]
which implies \begin{equation} F[N]=\int\limits_{U} {\rm Tr}[ N
{F}(A)]=\lim_{\epsilon\rightarrow 0}\ \sum_{p^i} {\rm
Tr}[N_{p^i}U_{p^i}[A]]\;,\end{equation} where the ${\rm Tr}$ in the 
right hand side is taken
in the fundamental representation. Notice that the explicit
dependence on the regulator $\epsilon$ has dropped out of the sum
on the r.h.s., a sign that we should be able to remove the
regulator upon quantization. With all this it is natural to write
the following formal expression for the generalized projection
operator, 
\begin{eqnarray} \label{twenty} 
{< P s,s^{\prime} >} & = & \lim_{\epsilon\rightarrow
0} \ \ <\prod_{p^i} \ \int \ dN_{p^i} \ {\rm exp}(i {\rm Tr}[
N_{p^i} \hat
  {U}_{p^i}]) s, \; s^{\prime}>\\  \nonumber
& = & \lim_{\epsilon\rightarrow 0} \  \ <\prod_{p^i} \
{\delta}(U_{p^i})s, \; s^{\prime}>,\label{P3} \end{eqnarray} 
where the last
equality follows from direct integration over $N_{p^i}$ at the
classical level; $\delta(U)$ being the distribution such that
$\int dg\ f(g) \delta(g)=f({1})$ for $f\in {\cal
L}^2(SU(2))$.
We can write $\delta(U)$ as a sum over unitary
irreducible representations of $SU(2)$, namely $
{\delta}({U_{p^i}})=\sum_j (2j+1) \ {{\chi}}_j(U_{p^i})$,
where ${\chi}_j(U)$ is the 
trace of the $j$-representation matrix of $U\in SU(2)$. It is  therefore
straightforward to evaluate explicitly the action of the projector.

We now proceed to construct the projector using our technique. We 
construct a Hamiltonian 
\begin{equation}
  H =\sum_{p} F^i_p F^i_p
\end{equation}
where $F^i_p \tau^i$ is the element of the $SU(2)$ algebra such that 
exponentiated gives the holonomy around the plaquette $p$ ($\tau^i$ are
the Pauli matrices), $U(p)=\exp(iF_p)$. We now proceed to construct the
projector as usual,
\begin{equation}
  {\cal P} = \lim_{\lambda\to\infty} \left({i\lambda \over \pi}\right)^{3/2}
e^{\left(-i\lambda H\right)}.
\end{equation}

If this expression is to agree with the projector computed by Noui and
Perez, one has to have that,
\begin{equation}
  \lim_{\lambda\to\infty} \int \left(\prod_{p} dg_l\right) f(g_{p_1},
g_{p_2},\ldots) {\cal P}_\lambda =
\int \left(\prod_{p} dg_l\right) f(g_{p_1},
g_{p_2},\ldots) \prod_p \delta(g_p).\label{84}
\end{equation}

In order to prove this identity we start by introducing a parameterization
of the $SU(2)$ group \cite{varshalovski},
\begin{equation}
  U = \left(
\begin{array}{cc}
a&b\\
-b^*&a^*
\end{array}\right),
\end{equation}
where 
\begin{eqnarray}
  a &=& = \cos\left({\omega \over 2}\right) 
- i \sin\left({\omega \over 2}\right)\cos\theta,\\
  b &=& = 
- i \sin\left({\omega \over 2}\right)\sin\theta e^{i\varphi}.
\end{eqnarray}
One then has that
\begin{equation}
  \int dg = 4 \int_0^\pi \sin^2\left({\omega \over 2}\right) d\omega \int_0^\pi \sin\theta d\theta\int_0^{2\pi}d\varphi,
\end{equation}
and
\begin{equation}
  \chi^j(\omega) = {\sin\left({(2j+1) \omega \over 2}\right) 
\over \sin\left({\omega \over 2}\right)}.
\end{equation}
And we see that $\omega=0$ corresponds to $a=1,b=0$. The $F$'s take the form,
\begin{eqnarray}
F^1&=&-\left({\omega \over 2}\right)\sin\theta\cos\varphi,\\
F^2&=& -\left({\omega \over 2}\right)\sin\theta\sin\varphi,\\
F^3&=&-\left({\omega \over 2}\right)\cos\theta.
\end{eqnarray}
From here we can compute $H$, 
\begin{equation}
  H= \sum_p \omega_p^2,
\end{equation}
and the projector can be written as,
\begin{equation}\label{94}
  {\cal P} = \lim_{\lambda\to\infty} \prod_p 
 \left( {i\lambda \over
  \pi} \right)^{3/2}
 e^{(-i\lambda \omega_p^2)}
\end{equation}
To complete the proof we would like to show that
\begin{equation}
  \delta(g_p) = {1 \over 4\pi} {d^2 \over d\omega^2} \delta(\omega) 
\equiv \delta(g_p(\omega,\theta,\varphi)),
\end{equation}
which can be shown by noticing that
\begin{eqnarray}
  \int dg_p f(g_p) \delta(g_p(\omega,\theta,\varphi)) &=& 2 \int_{-\pi}^\pi
d \omega\int d\Omega \sin^2\left({\omega \over 2}\right)f(g_p(\omega,
\theta,\varphi)) {1 \over 4\pi} {d^2 \over d\omega^2} \delta(\omega)\nonumber\\
&=&2 \int_{-\pi}^\pi d\omega\sin^2\left({\omega \over 2}\right)f(g_p(\omega,\theta,\varphi)) {d^2 \over d\omega^2}\delta(\omega)\nonumber\\
&=&2 \int_{-\pi}^\pi d\omega f(g_p(\omega,\theta,\varphi))\delta(\omega)
{d^2 \over d \omega^2}\sin^2 \left({\omega \over 2}\right)\nonumber\\
&=&  \int_{-\pi}^\pi d\omega f(g_p(\omega,\theta,\varphi))\delta(\omega)
\cos\omega=f({\leavevmode\hbox{\small1\kern-3.8pt\normalsize1}}).\label{99}
\end{eqnarray}
Finally, using the following Dirac identity for distributions,
\begin{equation}
  \sqrt{\frac{i \lambda}{\pi}}e^{-i\lambda \omega_p^2} = 
\sum_{j=0}^\infty \left({-i\over 2 \lambda}\right)^j { 1 \over (2j)!!} \delta^{(2j)}(\omega_p),
\end{equation}
where $\delta^{(2j)}$ is the $(2j)$-th derivative of the Dirac delta.
Inserting this expression in (\ref{94}) and using (\ref{99}) we 
prove the identity (\ref{84}) which implied that the projector we 
constructed with our technique agrees with the one obtained via
group averaging by Noui and Perez.

\section{Conclusions}

We have introduced a new approach to the quantization of totally
constrained systems and worked out several examples. One reaches two
conclusions. On one hand, one can use the technique to construct in a
well defined way the physical space of states of the totally
constrained theory of interest. Of special appeal is the fact that one
does it without the need of promoting the constraints to quantum
operators, which in several situations is known to be problematic. In
particular one does not have to deal with the problem of reproducing
the quantum constraint algebra.  In examples we have shown that the
results coincide with those of the group averaging procedure, where
the latter exists. Moreover one has a correspondence principle in that
one needs to reproduce the classical equations of motion as quantum
operatorial relations that offers guidance in the intermediate steps
of the process. The technique therefore has aspects that are quite
distinct from conventional ones, especially when one treats field
theories.

On the other hand the technique can be viewed as a new paradigm for
dealing with cases where the continuum theory does not exist. In those
cases one builds discrete theories where one can introduce a
relational notion of time and compute probabilities and expectation
values.  In examples we have shown that if one is able to take the
continuum limit, one reproduces predictions of the evolving constants
approach and therefore recovers all the physics in the theory. But we
have also shown in examples where the continuum limit does not exist
that one can define semiclassical regimes that reproduce, at certain
scales, the classical results of the theory of interest. Although we
do not yet know if the continuum limit exists in the case of general
relativity, this point of view would agree with popular beliefs in
that case: one would have a fundamental, discrete theory that
reproduces, at large scales, general relativity even if the quantum
theory strictly does not exist in the continuum limit.

Finally, the uniform discretizations also limit importantly the usual
ambiguities that appear when one discretizes theories and may even be
useful in the classical modeling of constrained theories.  Future
steps will require to test the technique in cases of increasing
complexity in the gravitational case. We have shown that in the
cosmological case one reproduces the results of loop quantum cosmology
and the technique may allow to explore some of the attractive features
of the latter in more complex models in the near future.

\section*{Acknowledgments}
We wish to thank Sebasti\'an Torterolo for help with the cosmology
section. This work was supported in part by grants NSF-PHY-0244335,
NSF-PHY-0554793, CCT-LSU, FQXi, Pedeciba, DID-USB grant GID-30 and Fonacit
grant G-2001000712 and the Horace Hearne Jr. Institute for Theoretical
Physics.

\end{document}